\documentclass[useAMS,usenatbib]{mn2e}

\usepackage{graphicx,amsmath,amssymb}
\usepackage{times}
\usepackage{color}
\usepackage{hyperref}

\newcommand{\ie}{i.e.,~}
\newcommand{\eg}{e.g.,~}

\title[Twisted-torus configurations of relativistic stars]{Twisted-torus
  configurations with large toroidal magnetic fields in relativistic
  stars }

\author[R. Ciolfi and L. Rezzolla]{R. Ciolfi and L.
Rezzolla\\
Max-Planck-Institut f\"ur Gravitationsphysik,
  Albert-Einstein-Institut, Potsdam, Germany}
\begin{document}



\maketitle

\label{firstpage}

\begin{abstract}
Understanding the properties of the internal magnetic field of neutron
stars remains a theoretical challenge. Over the last years, twisted-torus
geometries have been considered both in Newtonian and
general-relativistic equilibrium models, as they represent a
potentially good description of neutron star interiors. All of these
works have found an apparent intrinsic limitation to geometries that
are \emph{poloidal-field-dominated}, with a toroidal-to-poloidal
energy ratio inside the star that are $\lesssim 10\%$, unless surface
currents are included and magnetic fields are allowed to be
discontinuous. 
This limitation is in stark contrast with the general expectation that
much higher toroidal fields should be present in the stellar interior
and casts doubt about the stability and hence realism of these
configurations. We here discuss how to overcome this limitation
by adopting a new prescription for the azimuthal currents that leads to
magnetized equilibria where the toroidal-to-total magnetic-field energy
ratio can be as high as 90\%, thus including geometries that are
\emph{toroidal-field-dominated}. Moreover, our results show that
for a fixed exterior magnetic-field strength, a higher toroidal-field
energy implies a much higher total magnetic energy stored in the star,
with a potentially strong impact on the expected electromagnetic and
gravitational-wave emission from highly magnetized neutron stars. 
\end{abstract}



\section{Introduction}

Magnetic fields represent a key aspect of the physics and astrophysics of
neutron stars (NSs). Observational evidence points at very strong
external (polar) magnetic field strengths, up to $10^{13}$~G for ordinary
NSs and $10^{15}$~G for magnetars \citep{Duncan1992}, while the internal
fields might be even stronger. All the dynamical processes connected to
present observations of NSs are affected or even directly produced by
magnetic fields as, for example, the dipole radiation or the magnetar
flaring activity. Moreover, magnetically induced deformations of the
stellar structure can make rotating neutron stars potentially detectable
sources of gravitational waves (GWs) \citep{Bonazzola1996, Cutler2002,
  Frieben2012}. Both the GW and the electromagnetic emissions depend
sensitively on the amount of magnetic field energy stored in the NS and
on its geometrical distribution. Despite its great relevance,
observations have not yet provided direct constraints on the internal
magnetic-field configuration.

Understanding the properties of the internal magnetic field of NSs is a
theoretical challenge which dates back to the early work of
\citet{Chandrasekhar1953}. Since then, a number or analytical and
numerical studies have been devoted to the construction of equilibrium
models of magnetized NSs, at first considering the simple purely poloidal
and purely toroidal geometries. However, already from the analytical work
on nonrotating magnetized stars of \citet{Markey1973}, \citet{Wright1973}
and \citet{Tayler1973}, there has been growing evidence that these simple
geometries would suffer from the so-called Tayler (or kink) instability,
acting on Alfv\'en timescales, and that a long-lived magnetic field has
to consist of mixed poloidal-toroidal fields.  These perturbative
predictions have recently seen a number of confirmations through fully
nonlinear simulations in general relativity~\citep{Lasky2011, Ciolfi2011,
  Ciolfi2012, Kiuchi2011}. These studies also show that a significant
magnetic helicity is produced and that the system tends to an equipartion
between poloidal and toroidal fields~\citep{Ciolfi2012}. Unfortunately,
the prospects of detecting GWs produced by the instability are
pessimistic~\citep{Zink2012,Lasky2012,Ciolfi2012}.

Among the possible configurations with a mixed magnetic field, the
so-called \emph{twisted-torus} geometry has recently emerged as a good
candidate for NS interiors. It consists of an axisymmetric mixed field
where the poloidal component extends throughout the entire star and to
the exterior, while the toroidal one is confined inside the star, in the
torus-shaped region where the poloidal field lines are closed. An
important feature of this geometry is that the magnetic field is not
entirely confined to the interior of the star, as considered, \eg by
\citet{Ioka04}, \citet{Haskell2008}, \citet{DuezMathis2010} and
\citet{Yoshida2012}, thus in better agreement with the observational
evidence of external fields. Furthermore, in a twisted-torus
configuration both of the magnetic-field components are continuous at the
stellar surface. A discontinuity in the magnetic field would require the
inclusion of surface currents~\citep{Colaiuda2008}, for which there is no
obvious way to generate and sustain them. An important indication in
favour of the twisted-torus configuration comes from the work of
\citet{BraithwaiteNord2006}, where this geometry emerged as the final
outcome of the evolution of initial random fields in a nonrotating fluid
star. A twisted-torus magnetic field also appears natural in terms of the
poloidal and toroidal-field instabilities. The first one takes place in
the closed-line region and produces there a stabilizing toroidal
component, while the toroidal-field instability occurs near the symmetry
axis and produces there a poloidal field.

Over the last years, twisted-torus geometries were considered in
Newtonian \citep{Tomimura2005, Yoshida2006, Lander:2009, Lander2012,
  Glampedakis2012, Fujisawa2012} and general-relativistic frameworks
(\citealt{Ciolfi2009,Ciolfi2010}; hereafter Paper I and Paper II). In
spite the different approaches adopted, all the twisted-torus models
proposed so far have given rather similar results in terms of the
possible configuration of magnetic fields. The most important of these
results is the apparently unavoidable restriction to
\emph{poloidal-field-dominated} geometries, with an upper limit of $\sim
10\%$ for the toroidal-to-poloidal energy ratio in the star, unless
surface currents and discontinuous magnetic fields are
included~\citep{Fujisawa2013}.

This limitation is far more serious than it may appear. First, a much
higher toroidal-field content is expected from the formation scenario of
highly magnetized NSs, simply as a result of strong differential rotation
in the nascent NS~\citep{ThompsonDuncan1993,Bonanno:2003uw}. Second,
higher toroidal-field energies are needed when considering the
magneto-thermal evolution of magnetars, their bursting activity and the
pulse profiles \citep{Pons2011}. Finally, all evidence is that
poloidal-field-dominated geometries are unstable on Alfv\'en timescales
\citep{Braithwaite2009,Lander2012,Lasky2011, Ciolfi2011} and hence the
twisted-torus configurations considered so far may not be realistic.

In this Letter we overcome this limitation by adopting a new prescription
for the azimuthal currents that leads to more generic twisted-torus
configurations. In this way we construct a new sample of magnetized
equilibria where the toroidal-to-total magnetic-field energy ratio can be
as high as 90\%, thus including \emph{toroidal-field-dominated}
geometries. Moreover, we find that for a fixed exterior magnetic field
strength, a higher relative content of toroidal field energy implies a
much higher total magnetic energy in the star, with a potentially strong
impact on the expected electromagnetic and GW emission properties of
highly magnetized NSs.


\section[]{The Mathematical Model}

We consider axisymmetric equilibrium configurations of a non-rotating
magnetized NS, infinitely conducting and surrounded by vacuum, obtained
assuming that the magnetic field acts as a perturbation of a spherically
symmetric background star [see Paper I for a discussion on these
  assumptions and on the conditions under which a superfluid interior may
  not be considered~\citep{Lander2013}]. The background NS has a
(gravitational) mass of $M=1.4~M_\odot$ and is described as a barotropic
fluid with polytropic equation of state (EOS) $p=K\rho^{\Gamma}$, where
$p$ is the fluid pressure and $\rho$ the rest-mass density, with $K=100$
and $\Gamma=2$ (in units in which $c=G=M_\odot=1$). More sophisticated
EOSs could be considered (as done in Papers I and II) but a polytropic
choice is sufficient for our purposes and has the advantage that the
configurations obtained can be easily employed as initial data for
dynamical simulations. Magnetic fields can be considered as a
perturbation on the stellar structure as long as the magnetic energy is
much smaller than the binding energy and essentially if $\lesssim
10^{17}$ G, see \citet{Giacomazzo:2010} for a recent example. In
practice, we fix the field strength at the pole to $B_p=10^{15}$~G,
consistent with the observed order-of-magnitude of magnetars.

An equilibrium magnetic-field configuration is found by solving the
Grad-Shafranov equation for the unknown function represented by the
azimuthal component of the vector potential $A_\phi \equiv
\psi(r,\theta)$ (see Paper I and II for details)
\begin{align}
& -\frac{e^{-\lambda}}{4\pi}\left[\partial_r^2\psi + \frac{\partial_r\nu -
\partial_r\lambda}{2}\partial_r\psi \right] -
\frac{1}{4\pi r^2}\left[\partial_\theta^2\psi-\cot{\theta}\,
\partial_\theta\psi \right]  
\nonumber\\
& \hskip 2.5cm =
\bar{J}_{\phi} + 
F (\rho+2p) r^2\sin^2{\theta} 
\nonumber\\
& \hskip 2.5cm =
\frac{e^{-\nu}}{4\pi}\beta\frac{d\beta}{d\psi} + 
F (\rho+2p) r^2\sin^2{\theta} 
\,,
\label{GSeq}
\end{align}
where the rest-mass density $\rho$, the pressure $p$ and the metric
functions $\nu(r)$, $\lambda(r)$ are known from the background solution,
while $\beta(\psi)$ and $F(\psi)$ are two arbitrary functions, the former
one expressing the azimutal current $\bar{J}_{\phi}$. Once a solution is found
for $\psi$, the magnetic-field components can be computed from the curl
of the vector potential. As surface boundary conditions we impose that
$\psi$ and its derivatives match the exterior vacuum solution. We
simplify the problem by assuming that the magnetic field is dipolar and
thus that $\psi(r,\theta)=-a_1(r)\sin^2{\theta}$.

\begin{figure}
\centering
     \includegraphics[width=2.55cm,height=4.53cm]{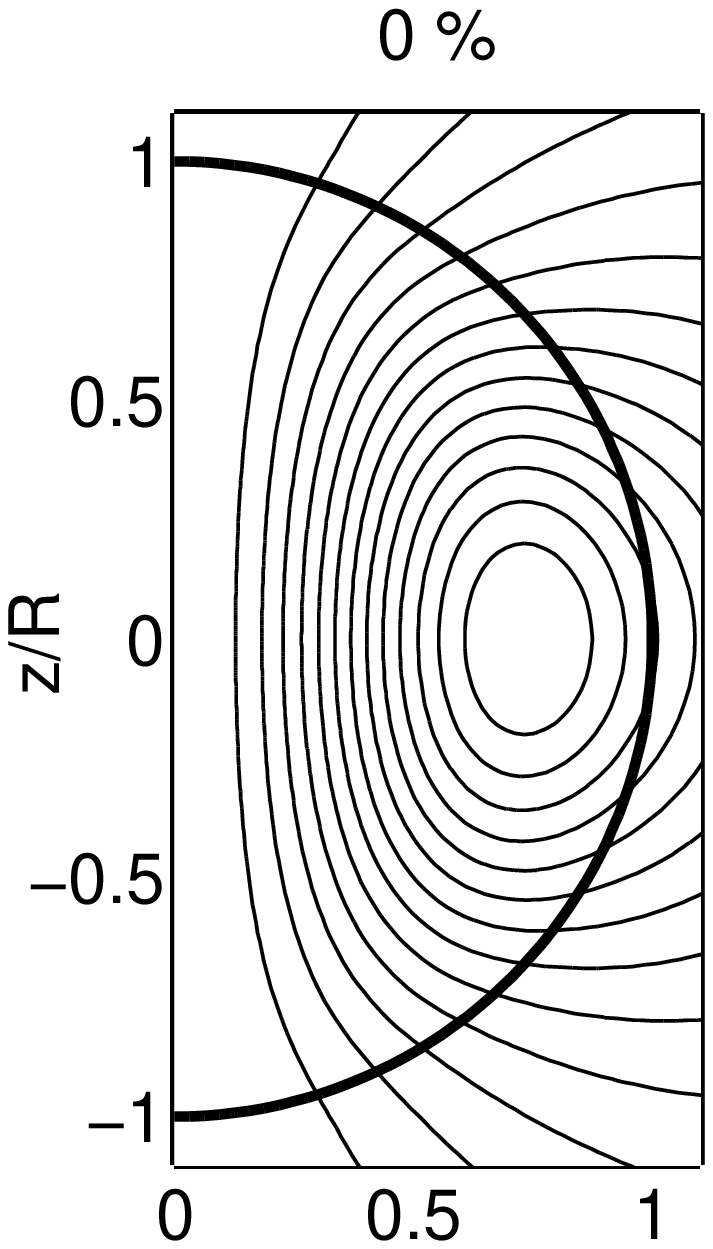}       
     \includegraphics[width=2.55cm,height=4.53cm]{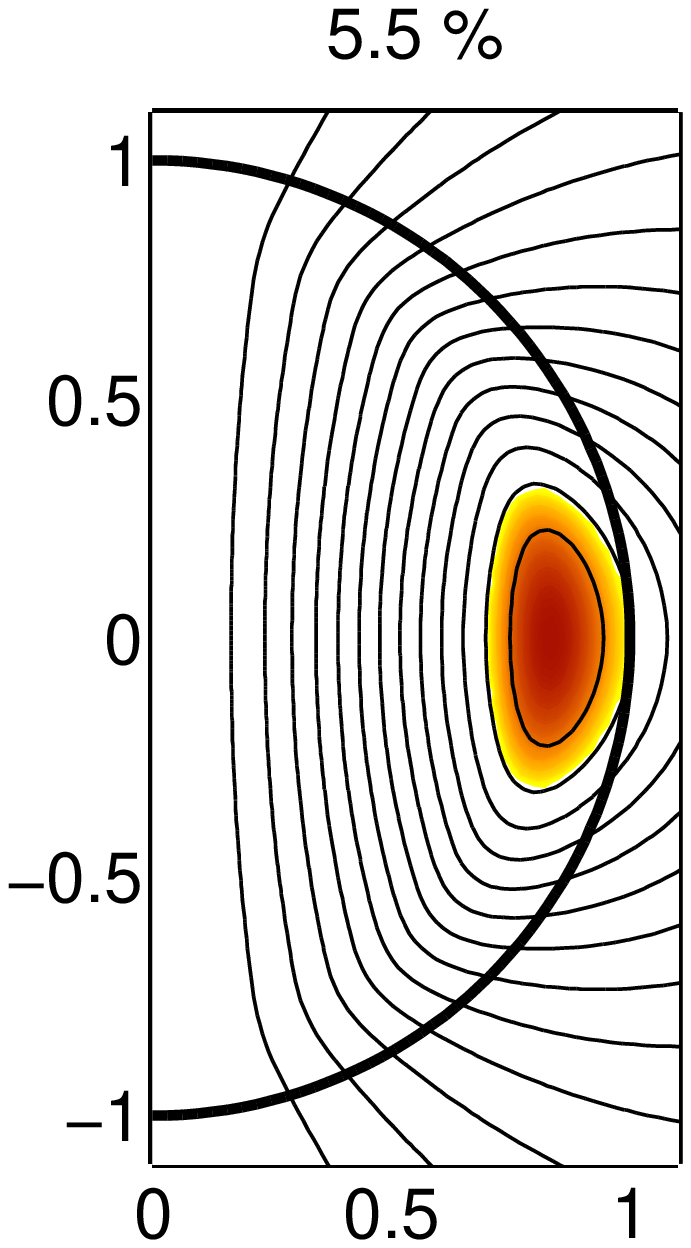}       
     \includegraphics[width=3.00cm,height=4.53cm]{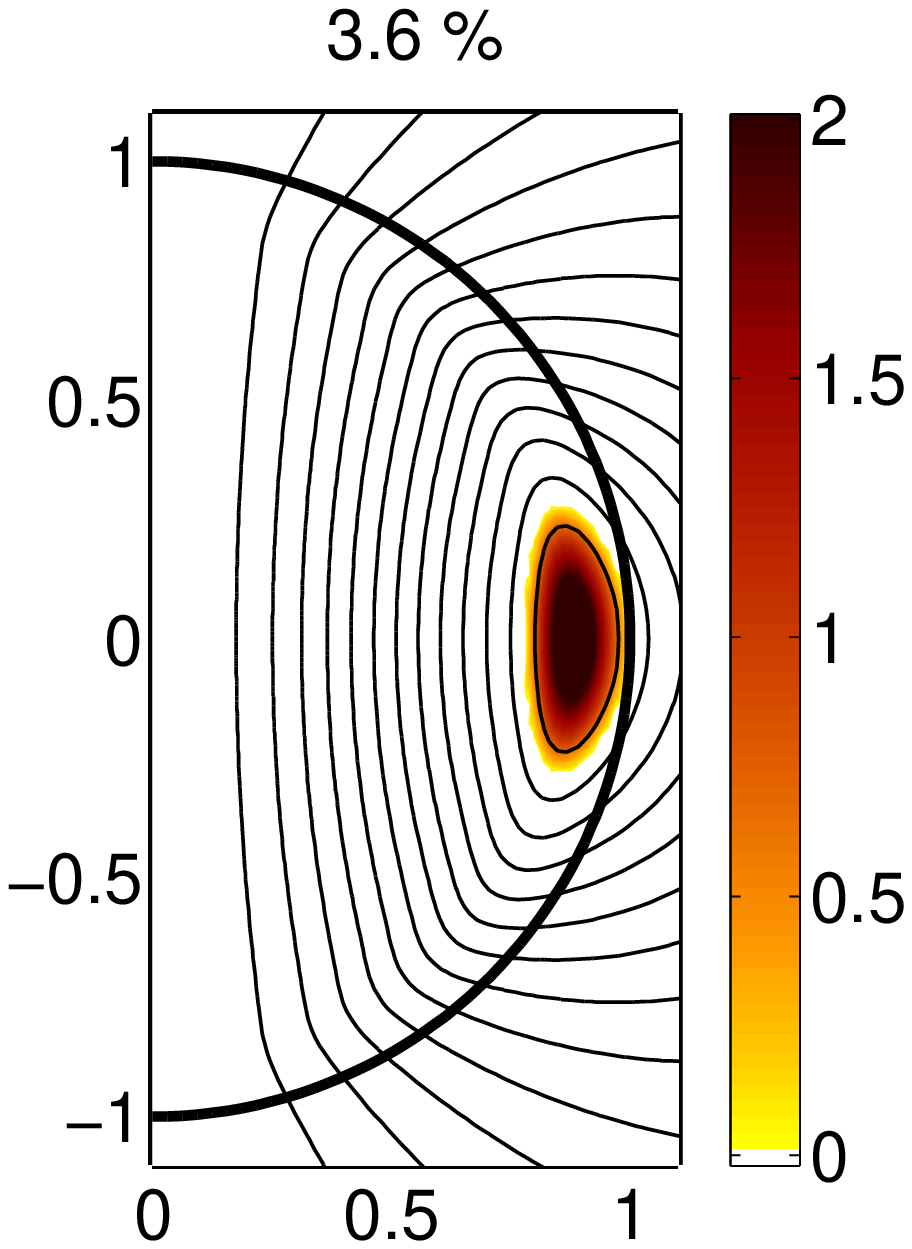}  
     \includegraphics[width=2.55cm,height=4.53cm]{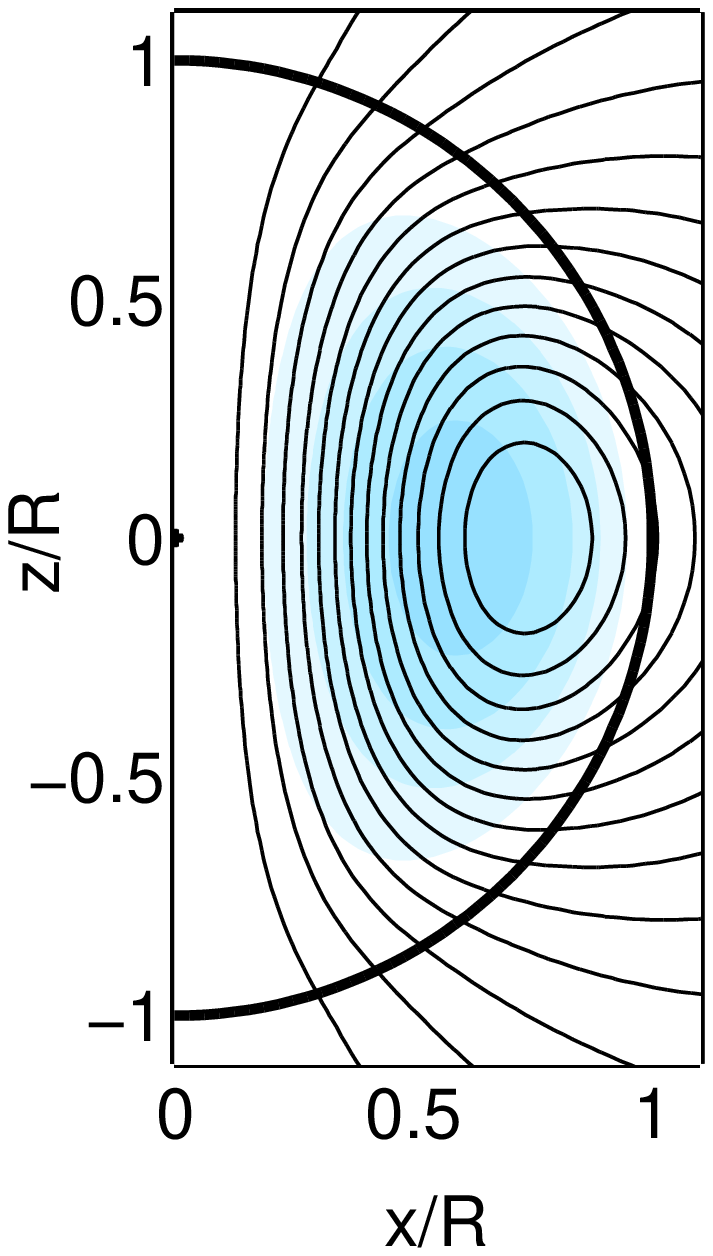}        
     \includegraphics[width=2.55cm,height=4.53cm]{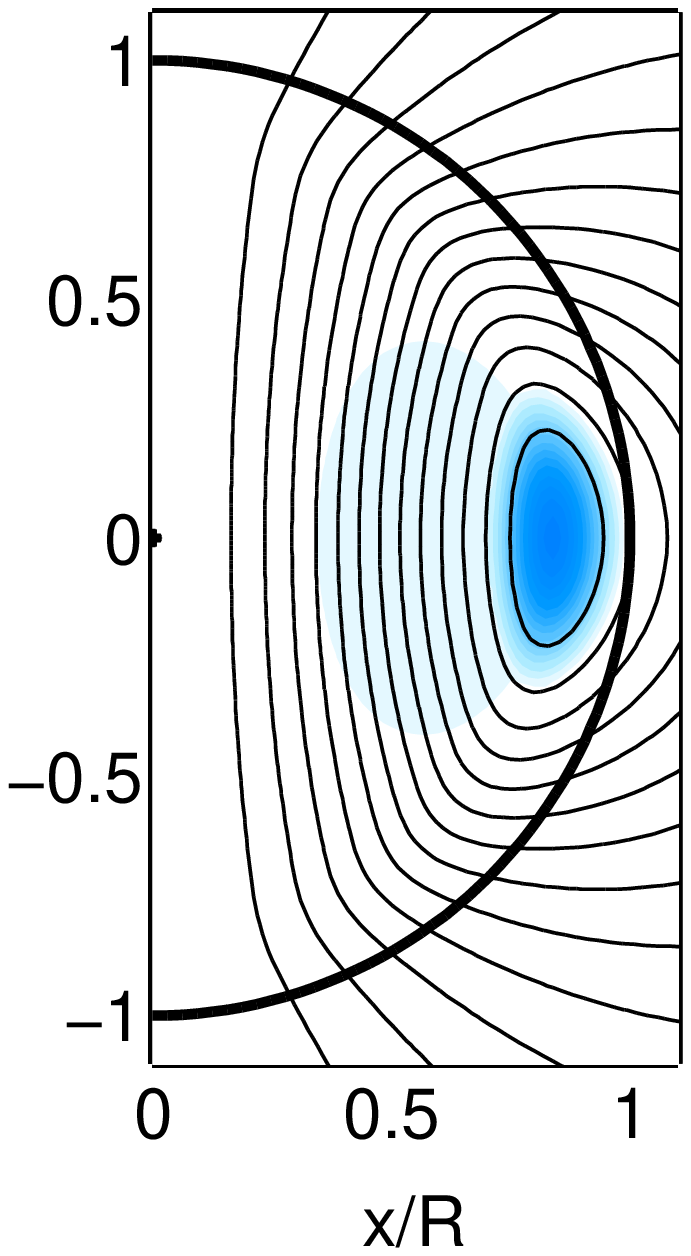}     
     \includegraphics[width=3.00cm,height=4.53cm]{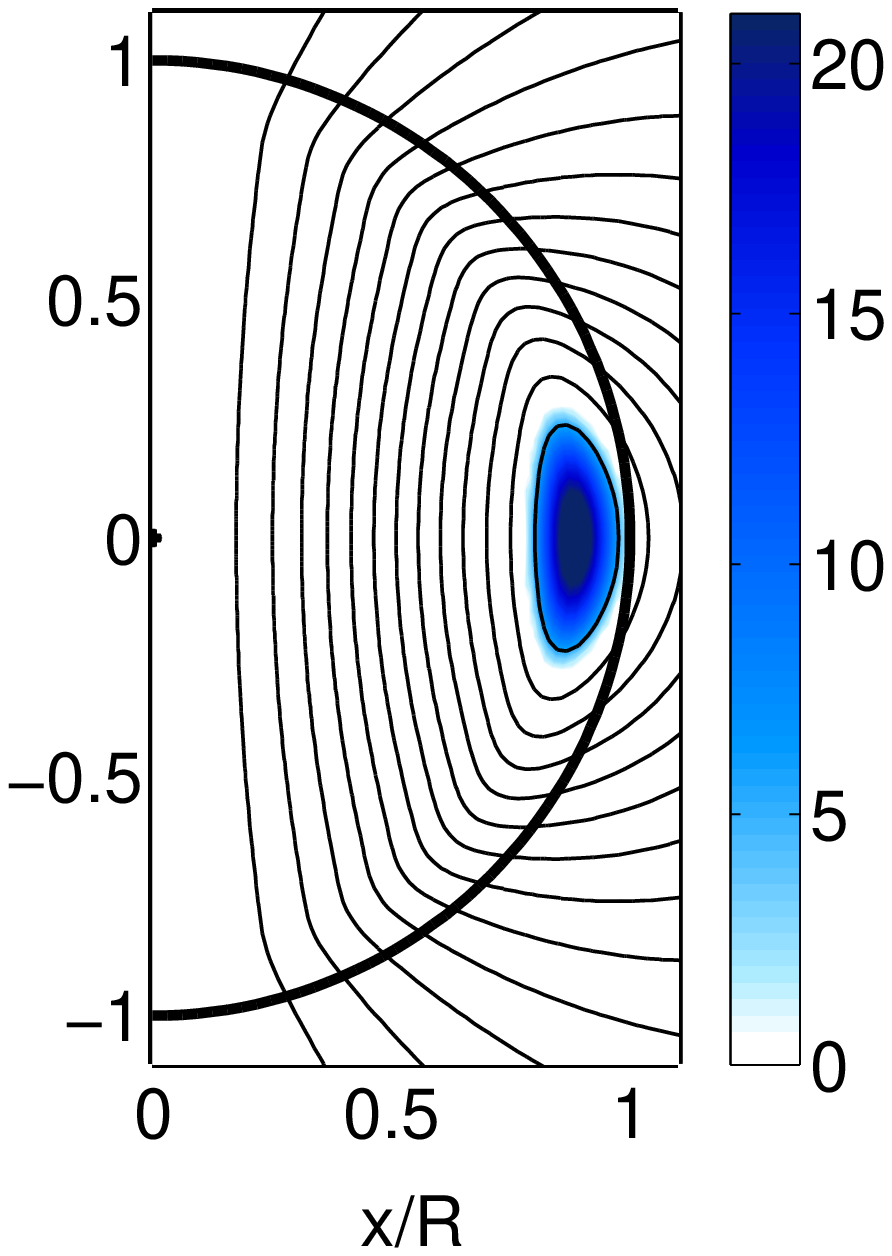}
  \caption{{\it Top panels:} Meridional view of magnetic field
    configurations obtained with constant $F(\psi)$. Shown in colour is
    the toroidal-field strength in units of the polar value $B_p=10^{15}$
    G. The labels on top of each panel show the toroidal-field energy
    content $E_{\mathrm{tor}} / E^{\,\mathrm{int}}_{\mathrm{m}}$. {\it
      Bottom panels:} The same as above but with a colour-coded
    representation of the azimuthal electric current $J_\phi$ in units of
    $10^{-5}$ km$^{-1}$. }
\label{fig:TT_oldJ}
\end{figure}

The choice of the trial functions $\beta(\psi)$ and $F(\psi)$ represents
a prescription for the current distribution and determines the final
magnetic field geometry. We recall that the function $\beta(\psi)$ is
related to the ratio of toroidal and poloidal fields (\ie $\beta=0$ gives
a purely poloidal field) and we take one of the simplest forms for a
twisted-torus geometry with continuous fields, $\beta(\psi) =
\zeta_0\psi\,(|\psi / \bar{\psi}|-1)\,\Theta(|\psi/\bar{\psi}|-1)$, where
$\zeta_0$ is a constant, $\bar{\psi}$ is the value of $\psi$ on the last
closed-field line (the one tangent to the surface) and $\Theta(x)$ is the
Heaviside step function, which confines the toroidal field in the
closed-line region. Once the second arbitrary function $F(\psi)$ is also
chosen, we can vary the constant $\zeta_0$ to obtain a set of
configurations with varying toroidal-field content. The simplest and
common choice for $F$ is that it is a constant, \ie
\begin{equation}
\label{eq:fpsi_const}
F(\psi)=c_0\,, 
\end{equation}
which then leads to the configurations shown in
Fig.~\ref{fig:TT_oldJ}. The initial solution has $\zeta_0=0$ (\ie no
toroidal field; left panels) and we increase $\zeta_0$ to reach a higher
internal toroidal-to-total magnetic-field energy ratio $E_{\mathrm{tor}}
/ E^{\,\mathrm{int}}_{\mathrm{m}}$. As a result, the toroidal field
becomes stronger, but the closed-line region also shrinks, so that the
amount of toroidal magnetic energy reaches a maximum with
$E_{\mathrm{tor}} / E^{\,\mathrm{int}}_{\mathrm{m}}=5.5\,\%$ (middle
panels) and then starts decreasing (right panels). This happens because
the stronger toroidal fields require a stronger electric current in the
closed-line region (Fig.~\ref{fig:TT_oldJ}, bottom panels) and this
affects the poloidal field lines by moving the neutral point outwards. In
practice, these examples summarize the limitations of the previous
twisted-torus models, which could lead to NSs with only $E_{\mathrm{tor}}
/ E^{\,\mathrm{int}}_{\mathrm{m}}\lesssim 10\%$. In what follows we
discuss how we can overcome these restrictions with a more ingenious
choice for $F(\psi)$. For simplicity, we maintain the same form for
$\beta(\psi)$ as different choices yield similar results (see Paper II).

The first element of our prescription to increase the amount of
toroidal-field energy in the star consists in having a larger region of
closed field lines. In absence of toroidal fields, the closed-line region
can be enlarged by changing the prescription for $F(\psi)$ so that the
electric currents are more concentrated near the symmetry axis. Note that
this also implies an increase of the magnetic energy in the star for a
fixed external field strength. Pushing this idea to the limit would give
a poloidal field entirely confined in the star (see also
\citealt{Fujisawa2012}). If toroidal fields are included, a larger
closed-line region would still undergo a contraction, but we expect the
maximum toroidal-field energy to be considerably larger. To produce
a larger region of closed field lines we extend~\eqref{eq:fpsi_const} as
\begin{equation}
F(\psi)=c_0 \left[
( 1-|\psi/\bar{\psi}| )^4 \Theta( 1-|\psi/\bar{\psi}|)-\bar{k}
\right]\,,
\label{Fofpsi}
\end{equation}
with $c_0$ and $\bar{k}$ constants. 

The second element of our prescription aims instead at reducing the
effect that toroidal fields have on poloidal field lines by modifying
\eqref{Fofpsi} as $F(\psi)\rightarrow F(\psi) + \bar{F}(\psi)$, with the
new term $\bar{F}(\psi)$ chosen so as to cancel as much as possible the
toroidal-field contribution to the Grad-Shafranov equation, \ie the
$\beta$ term in Eq.~(\ref{GSeq}). In practice, we set $\bar{F} =
X(\psi)\, \beta\, ({d\beta}/{d\psi})$, so that the complete azimuthal
current that we want to minimize is
\begin{align}
\hat{J}_{\phi} & \equiv \bar{J}_{\phi}+\bar{F}(\rho+2p) r^2\sin^2{\theta}   
\nonumber\\
& = \beta\frac{d\beta}{d\psi}
\left[\frac{e^{-\nu}}{4\pi}+ X(\psi) (\rho+2p)r^2\sin^2{\theta} \right]\,.
\end{align}
Making this quantity negligibly small would allow us to add an
arbitrarily large toroidal field without changing the closed-line region
and thus resulting in an arbitrarily high $E_{\mathrm{tor}} /
E^{\,\mathrm{int}}_{\mathrm{m}}$. Despite our freedom in choosing
$X(\psi)$, a perfect cancellation would be possible only in the limit of
a magnetic field entirely confined inside the star. Nevertheless, the
minimization of $\hat{J}_{\phi}$ becomes more effective as the
closed-line region extends towards the symmetry axis and at some point it
starts increasing significantly the maximum toroidal-field energy. Again,
the possibility of storing more energy in toroidal fields also leads to
the increase of the total magnetic energy in the star. Here we
simply set $X(\psi)=X_0 = \textrm{const.}$, with $X_0$ fixed so as to
minimize $\hat{J}_{\phi}$. Remarkably, prescription~(\ref{Fofpsi}) and
the addition of $\bar{F}$ have a limited effect if adopted
separately, but their combined use yields a significant difference.

\begin{figure}
\centering
      \includegraphics[angle=0,height=4.5cm]{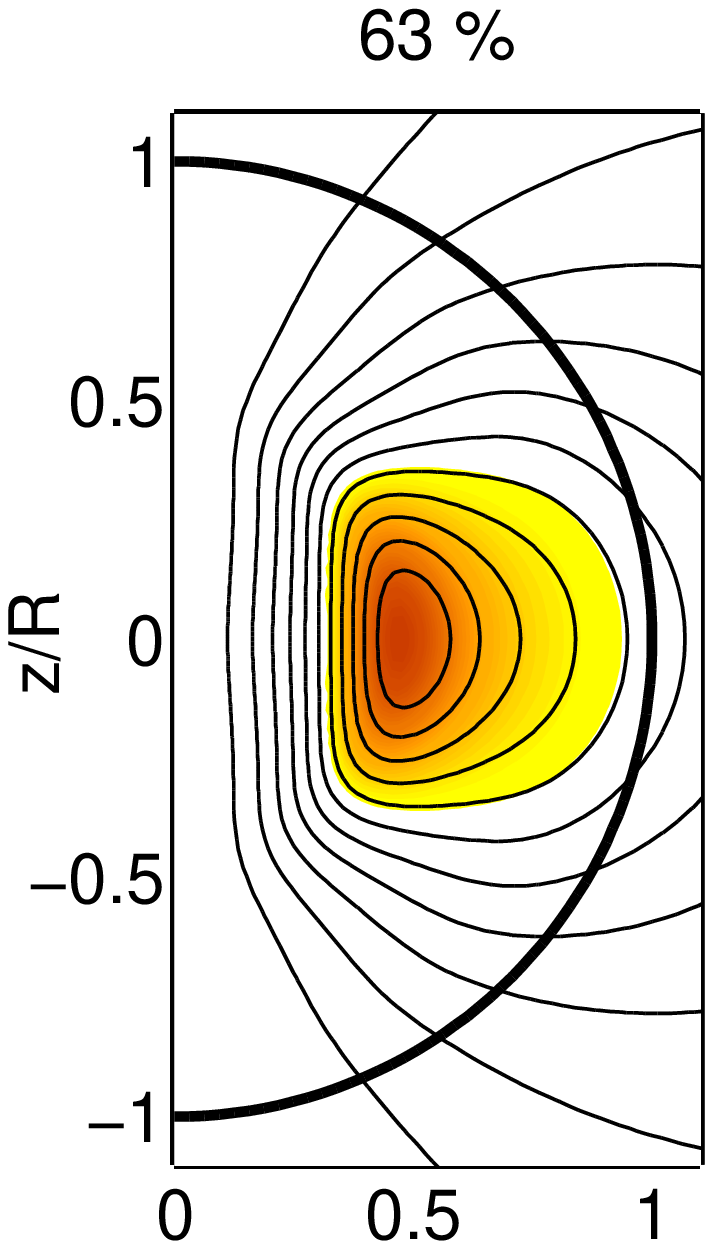}
      \includegraphics[angle=0,height=4.5cm]{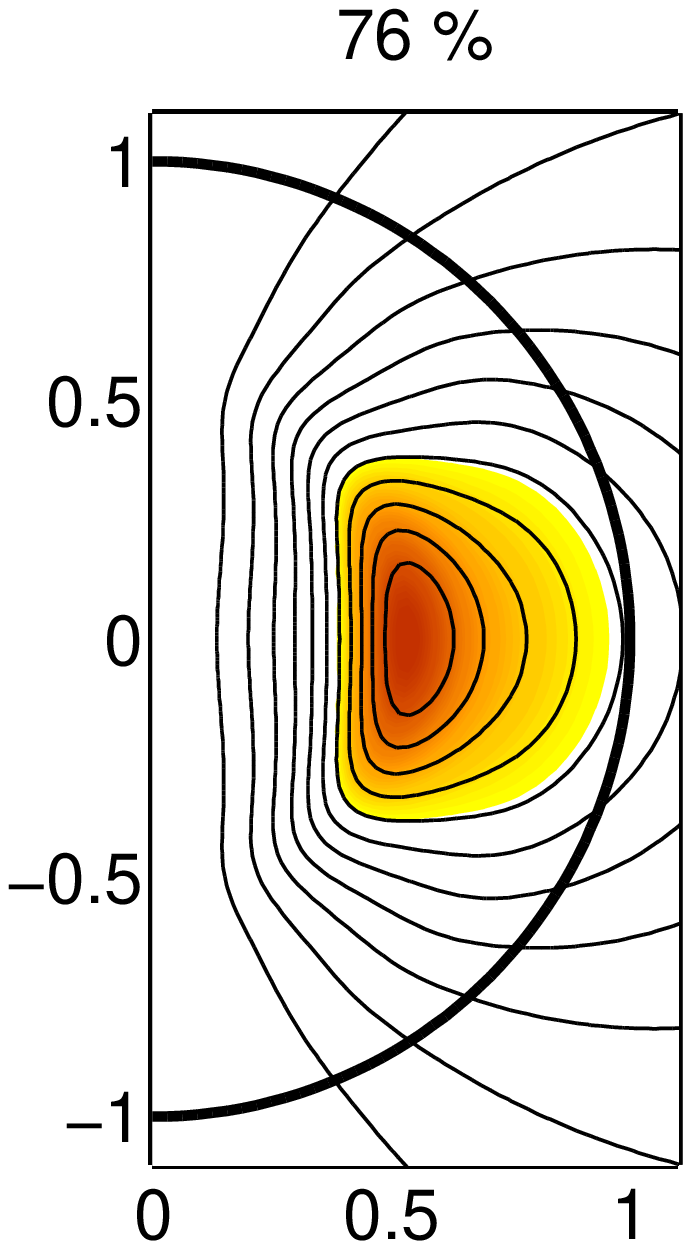}
      \includegraphics[angle=0,height=4.5cm]{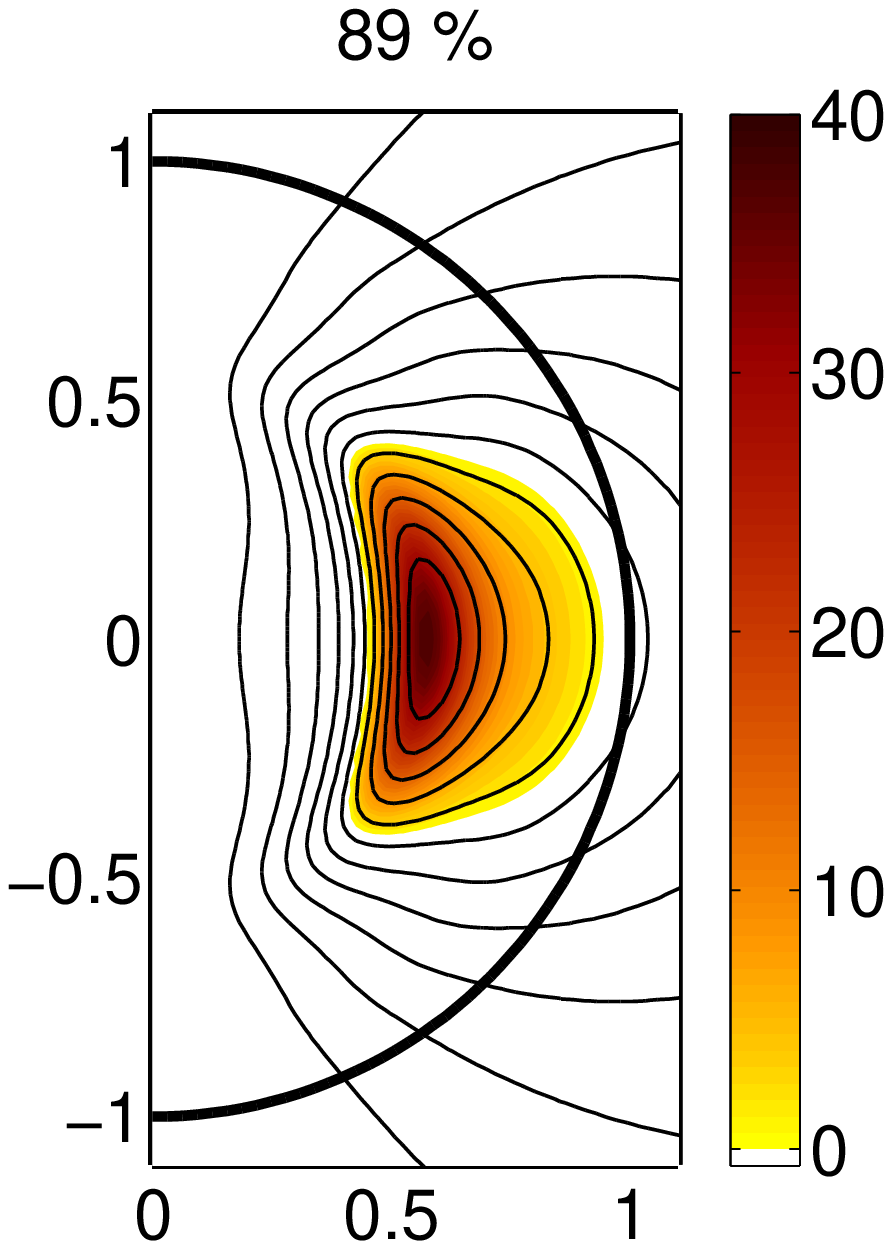}
      \hglue 0.05cm
      \includegraphics[angle=0,height=4.5cm]{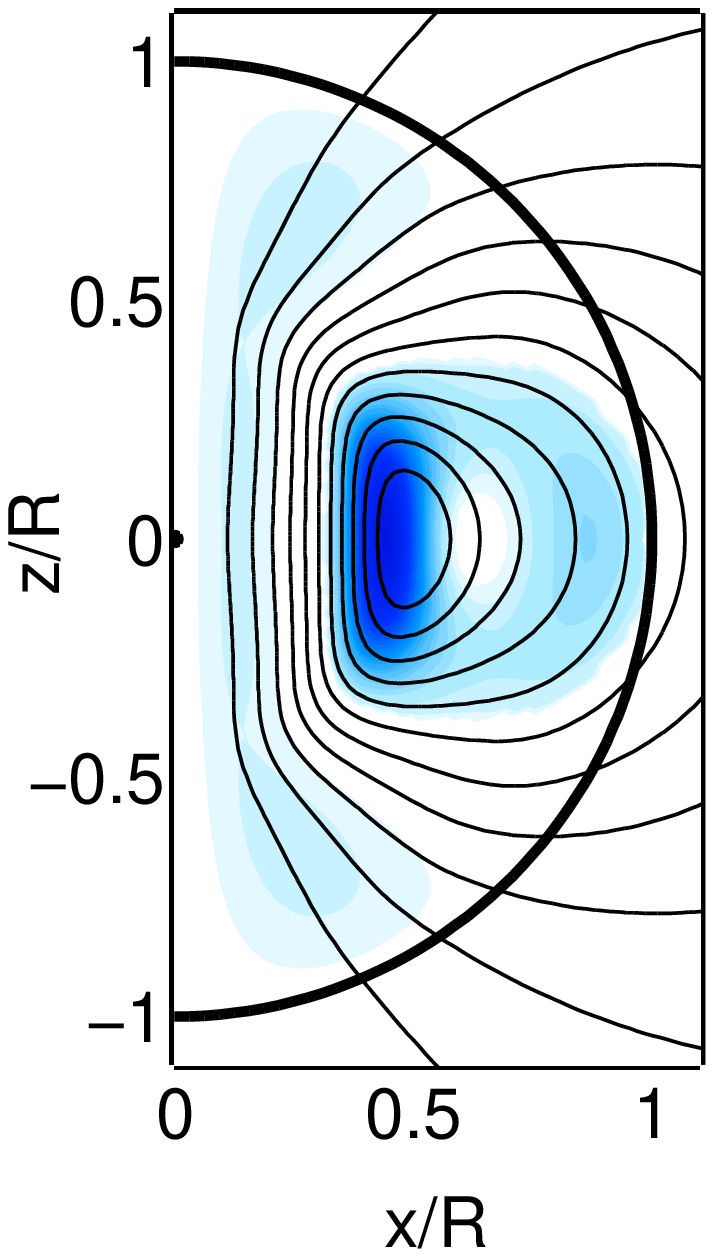}   
      \includegraphics[angle=0,height=4.5cm]{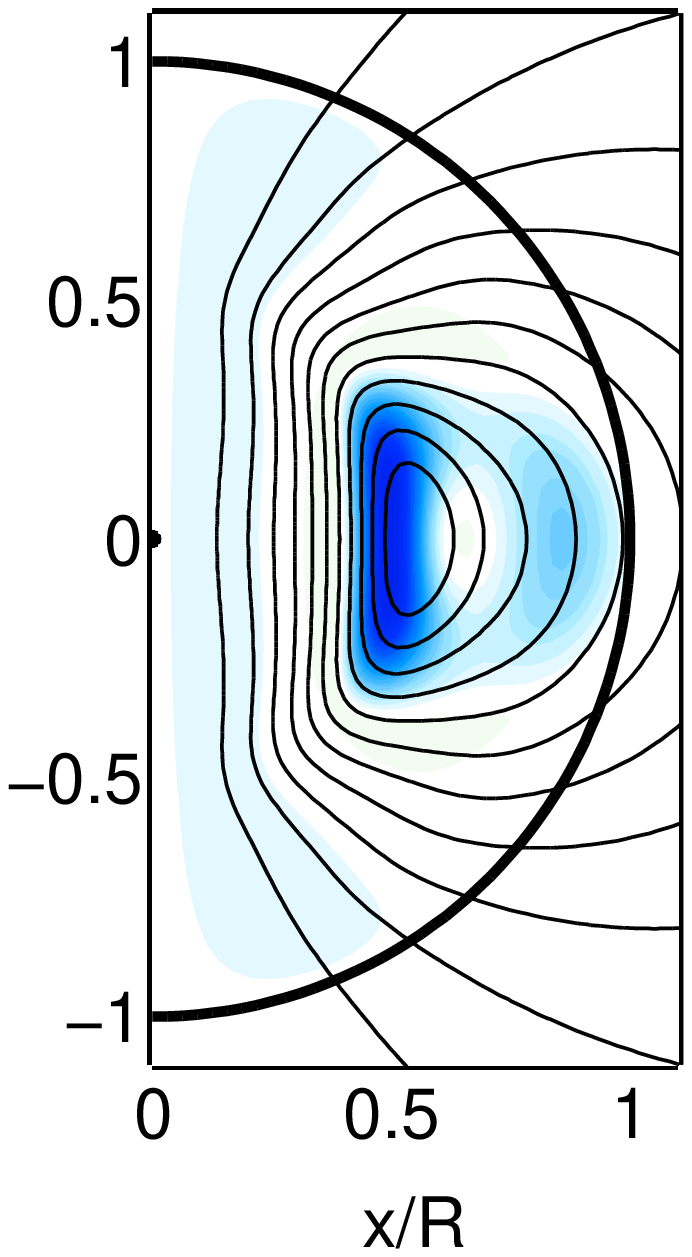}
      \includegraphics[angle=0,height=4.5cm]{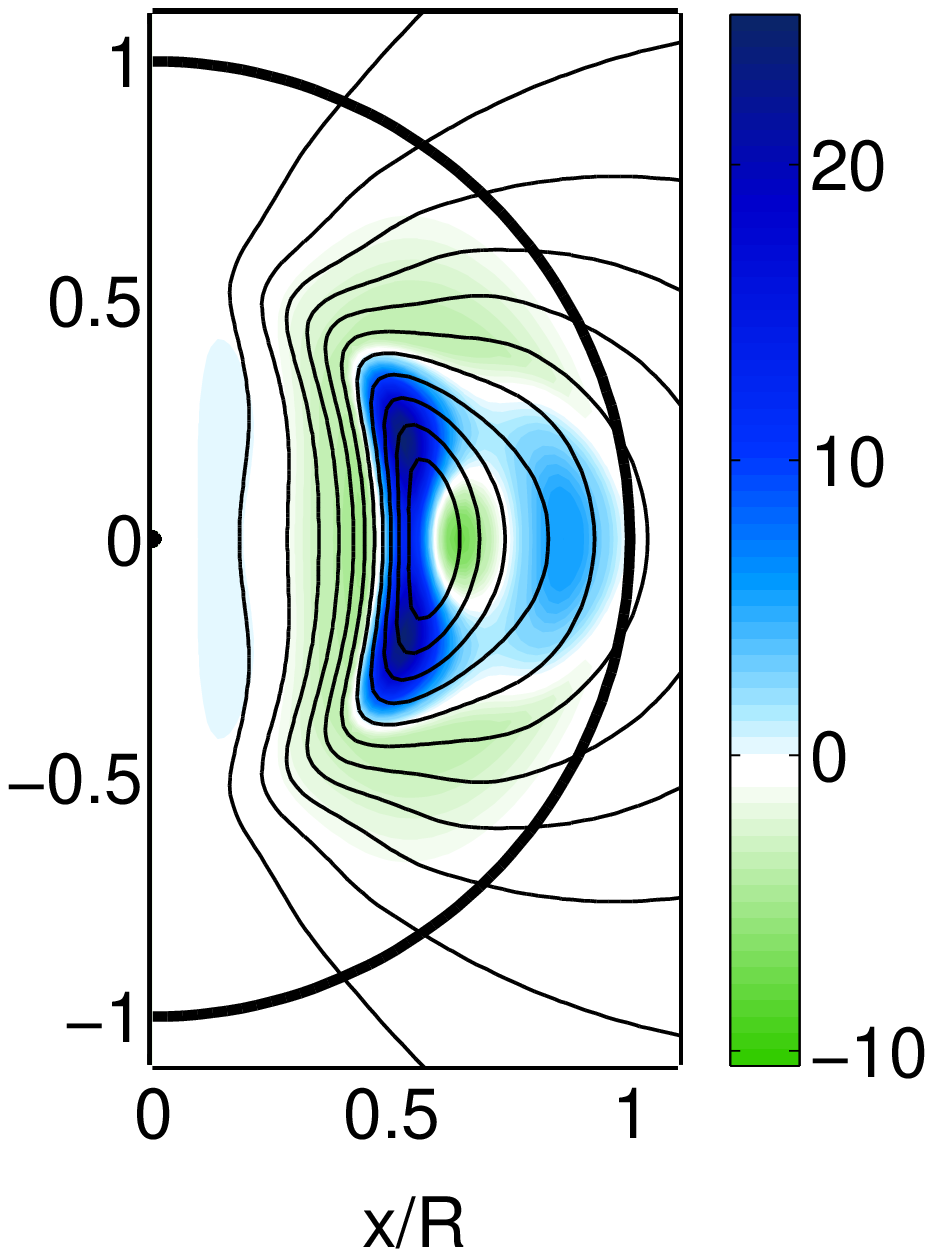}
     \caption{The same as in Fig.~\ref{fig:TT_oldJ} for three
       configurations of maximal $E_{\mathrm{tor}} /
       E^{\,\mathrm{int}}_{\mathrm{m}}$ obtained with the new
       prescription; from left to right $\bar{k}= 0.03, 0.15, 0.35$. }
\label{fig:TT_newJ}
\end{figure}


\section[]{New Twisted-torus solutions} 

Using our new approach we consider different example models with
$\bar{k}$ in the range $[0.03,0.35]$. For each $\bar{k}$, we span the
full range of $\zeta_0$, obtaining geometries with varying
$E_{\mathrm{tor}} / E^{\,\mathrm{int}}_{\mathrm{m}}$. All the other
constants are fixed once we choose $B_p=10^{15}$~G. Using the same
conventions as in Fig.~\ref{fig:TT_oldJ}, we show in
Fig.~\ref{fig:TT_newJ} the configurations having the maximum toroidal
field content for three choices of $\bar{k}$, noting that in general a
higher $\bar{k}$ gives a higher energy ratio, up to $E_{\mathrm{tor}} /
E^{\,\mathrm{int}}_{\mathrm{m}} = 89\%$. This extreme case corresponds to
a toroidal-field-dominated geometry and clearly demonstrates that the new
prescription allows us to overcome the limitations of previous
twisted-torus models. Note also from the different colour scale that the
new configurations have much stronger maximum toroidal fields
$B^{\mathrm{max}}_{\mathrm{tor}}$, up to $\sim 10^{16}$ G, while the
currents are comparable.

\begin{figure*}
\centering
\begin{minipage}{176mm}
  \begin{center}
     \includegraphics[angle=0,height=5.8cm]{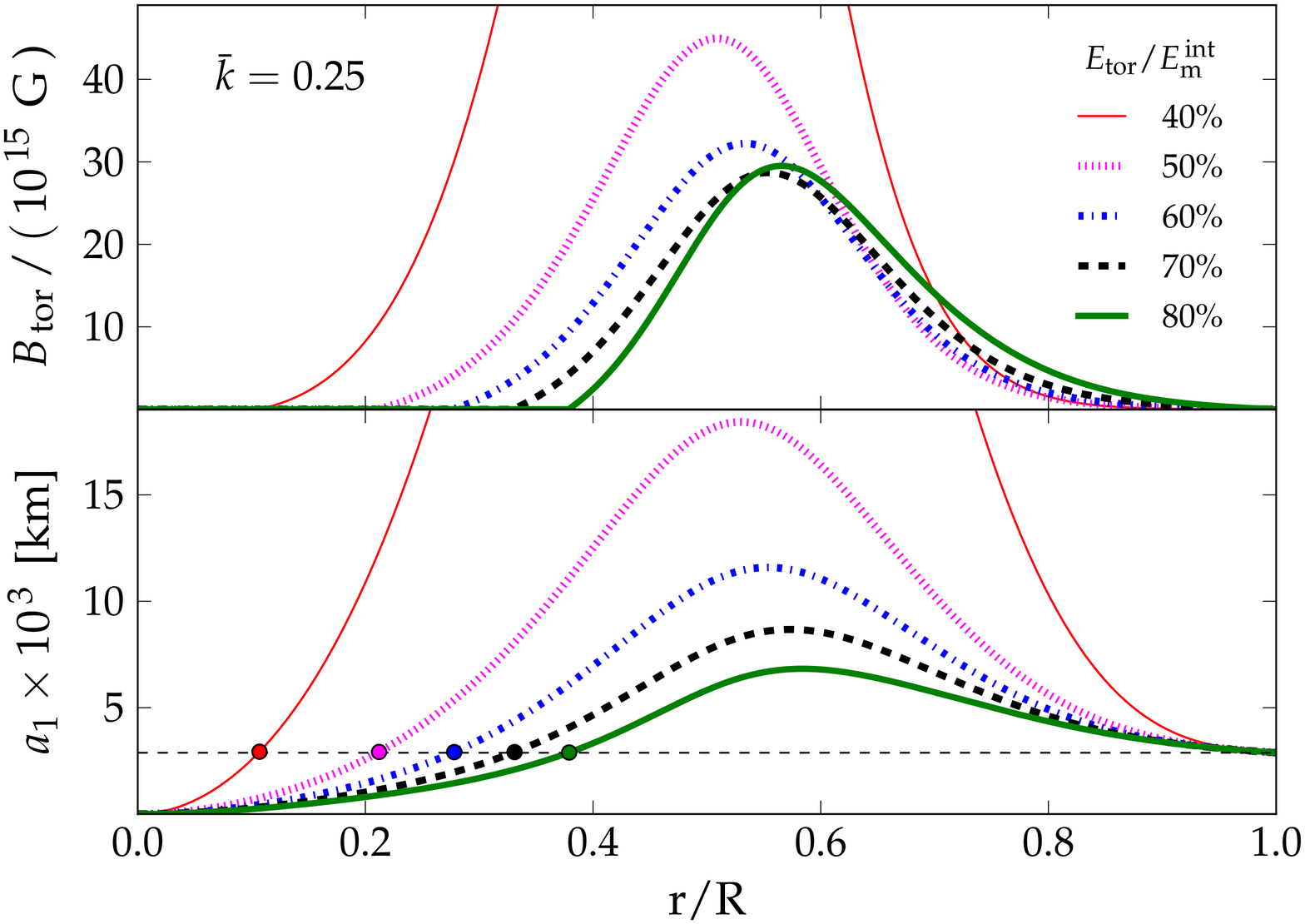}
\hspace{0.8cm}
     \includegraphics[angle=0,height=5.8cm]{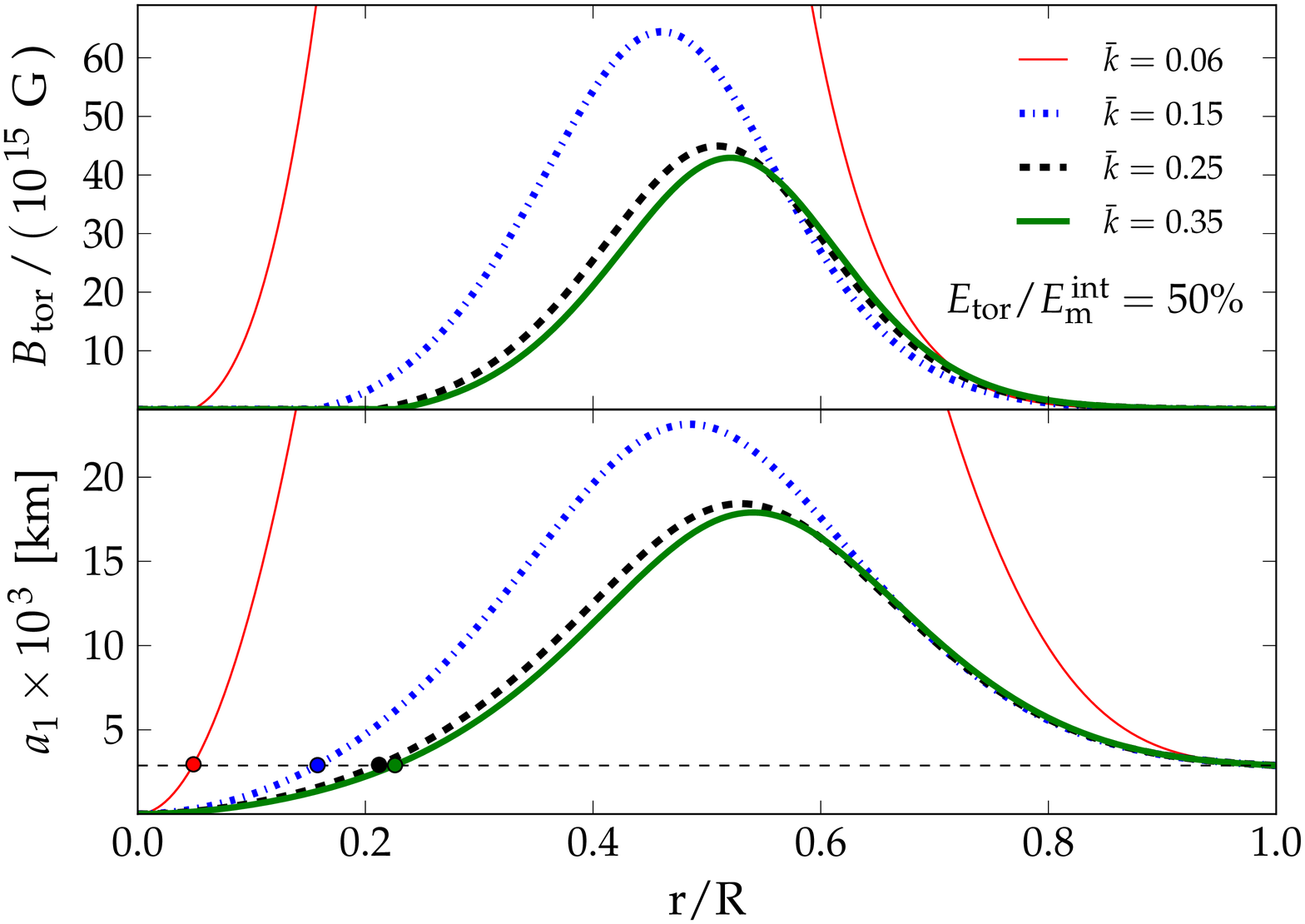}
  \end{center} 
  \caption{{\it Left panel:} The top part shows the radial profile of the
    toroidal magnetic field on the equatorial plane for models with
    $\bar{k}=0.25$ and different toroidal-field content $E_{\mathrm{tor}}
    / E^{\,\mathrm{int}}_{\mathrm{m}}$. The bottom part, instead, shows
    the radial profile of $a_1(r)=-\psi(r,\theta=\pi/2)$ for the same
    models, with the filled dots marking the extension of the
    closed-line region. {\it Right panel:} The same as in the left
    panel, but for models with fixed $E_{\mathrm{tor}} /
    E^{\,\mathrm{int}}_{\mathrm{m}}$ and varying $\bar{k}$. }
\label{fig:1Dplots}
\end{minipage}
\end{figure*}

Figure~\ref{fig:1Dplots} provides a more quantitative measure of the
magnetic field and the vector potential via the function $a_1(r)$, with
the left panel illustrating configurations obtained with $\bar{k}=0.25$
and increasing $E_{\mathrm{tor}} / E^{\,\mathrm{int}}_{\mathrm{m}}$, from
40\% to 80\% (the top part shows the radial profile of the toroidal
magnetic field on the equatorial plane, while the bottom part shows the
radial profile of $a_1(r)$ for the same models). Note that for fixed
$\bar{k}$, a higher toroidal-field content reduces the extension of the
closed-line region. On the other hand, as this region expands when
keeping fixed the polar field strength $B_p$, both the interior poloidal
and toroidal-field strengths increase very rapidly. In the
right panel of Fig.~\ref{fig:1Dplots} we report the same quantities as in
the left one but when varying $\bar{k}$ and imposing $E_{\mathrm{tor}} /
E^{\,\mathrm{int}}_{\mathrm{m}}=50\%$, \ie the energy equipartition
between poloidal and toroidal fields. As the sequence shows, when
approaching $\bar{k}=0.35$, the strength of the two components rapidly
converges to a minimum value. In terms of the internal magnetic energy,
this lower limit gives $E^{\,\mathrm{int}}_{\mathrm{m}} / E^*\gtrsim 68$,
where $E^*\simeq 2.7\times10^{48}$~erg is the internal magnetic energy of
the purely poloidal model shown in the first panel of
Fig.~\ref{fig:TT_oldJ}. We conclude that for $E_{\mathrm{tor}} /
E^{\,\mathrm{int}}_{\mathrm{m}}=50\%$, the internal magnetic energy is at
least 1-2 orders of magnitude larger than for the
poloidal-field-dominated models built with the previous prescription
(\ref{eq:fpsi_const}) and the same $B_p$. This result provides an
effective example of how a higher toroidal field content implies a much
higher internal magnetic energy.

Although our findings depend quantitatively on the particular choice for
$F(\psi)$ and $\beta(\psi)$, they imply the following result that we
expect to be general for twisted-torus geometries: if a more substantial
part of magnetic energy is in the toroidal-field component, \ie if
$E_{\mathrm{tor}} / E^{\,\mathrm{int}}_{\mathrm{m}} \gtrsim 10\%$, then
higher internal magnetic energies are not only possible but rather
inevitable.


\section[]{Quadrupolar Deformations}

Magnetic fields alter the NS density and pressure distributions inducing
quadrupolar deformations that can be quantified in terms of the
quadrupolar ellipticity $\epsilon_{_{Q}} \equiv Q/I$, where $Q$ is the
mass-energy quadrupole moment and scales as the magnetic energy (\ie
$\propto B^2$), while $I$ is the mean value of the stellar moment of
inertia [see also the discussion by~\citet{Frieben2012} on the difference
  between surface and quadrupolar ellipticities]. Poloidal fields deform
a nonrotating star towards an oblate shape (equatorial radius larger than
the polar radius), corresponding to positive $\epsilon_{_{Q}}$, whereas
toroidal fields have the opposite effect; therefore, the amount of
deformation is reflected in the toroidal-to-poloidal energy ratio. In
general, a magnetized NS that rotates around an axis misaligned with
respect to the magnetic axis and having $\epsilon_{_{Q}} \neq 0$, will
emit a continuous GW signal with amplitude $h \propto |\epsilon_{_{Q}}|
I\Omega^2/d$ \citep{Bonazzola1996}, where $\Omega$ is the angular
velocity and $d$ the source distance from the observer.  Following the
procedure adopted in Paper II, we can compute the deformations of our new
set of configurations and compare them with the typical predictions of
previous twisted-torus models. Considering a set of realistic EOSs would
introduce a variance of $\sim 2$ in the results (see Paper II); however,
here we are concerned with the influence that the magnetic field geometry
has on the deformation, which is independent of the particular EOS
adopted.

In Table~\ref{tab1} we report the results for some representative
solutions, with the first two lines referring to configurations obtained
with prescription (\ref{eq:fpsi_const}), corresponding to the purely
poloidal model and to one with maximal toroidal energy (first two columns
in Fig.~\ref{fig:TT_oldJ}). These models are representative of the
order-of-magnitude deformations obtained with the
poloidal-field-dominated solutions proposed so far in the literature. The
other lines refer to our new configurations with different values of
$\bar{k}$ and toroidal magnetic energies. In addition to
$\epsilon_{_{Q}}$ for $B_p=10^{15}$~G, we also show its value normalized
to the internal magnetic energy $E^{\,\mathrm{int}}_{\mathrm{m}}$
(expressed in km). This quantity is independent of the magnetic-field
strength and only depends on the geometrical distribution.

As mentioned above, the sign of $\epsilon_{_{Q}}$ results from the
balance between poloidal (positive) and toroidal (negative) field
deformations and the separation occurs around equipartition, \ie
$E_{\mathrm{tor}} / E^{\,\mathrm{int}}_{\mathrm{m}} \sim 50\%$,
indicating that such balance is essentially controlled by the amount of
energy in the two components. The same can be deduced by looking at the
normalized ellipticities
$\epsilon_{_{Q}}/E^{\,\mathrm{int}}_{\mathrm{m}}$. These are smaller when
$ E_{\mathrm{tor}} / E^{\,\mathrm{int}}_{\mathrm{m}} \simeq 40-50\%$,
where the two kind of deformations are well balanced, and they grow when
moving away from equipartition. In particular, similar (and opposite)
values are found in the poloidal-field-dominated and in the
toroidal-field-dominated case. Note that different values of $\bar{k}$
with the same toroidal energy content give only slightly different
$\epsilon_{_{Q}} / E^{\,\mathrm{int}}_{\mathrm{m}}$. Most importantly,
the majority of the new configurations considered has
$|\epsilon_{_{Q}}|\sim 10^{-4}\times(B_p/10^{15}~\text{G})^2$, which is
at least \emph{one order of magnitude} larger than the values given by
poloidal-field-dominated geometries. This is due to the much higher
internal magnetic energies obtained for geometries with higher
$E_{\mathrm{tor}} / E^{\,\mathrm{int}}_{\mathrm{m}}$. In summary,
twisted-torus configurations with higher toroidal energy content, say,
$E_{\mathrm{tor}} / E^{\,\mathrm{int}}_{\mathrm{m}}\geq 40\%$, can easily
produce GW amplitudes that are one order of magnitude larger than what
predicted so far, either in newly-born magnetars or in pulsars with large
buried fields, with an obvious enhancement of their
detectability. Moreover, in magnetized NSs with negative ellipticities, a
spin-flip mechanism driven by viscosity may occur, leading to an increase
in the angle between the spin and magnetic axes, which tend to become
nearly orthogonal \citep{Jones1975,Cutler2002}. In this case, the GW
emission would be further enhanced, with optimistic prospects of
detection \citep{Stella2005}.

\begin{table}
  \begin{center}
  \begin{tabular}{cccrr}
\hline
\hline
$\bar{k}$ & $E_{\mathrm{tor}} / E^{\,\mathrm{int}}_{\mathrm{m}}$ &
$\epsilon_{_{Q}}$ & 
$\epsilon_{_{Q}}/E^{\,\mathrm{int}}_{\mathrm{m}}\,[\text{km}^{-1}]$ & $B_\mathrm{tor}^\mathrm{max}/B_p$ \\
\hline
$-   $   \!\!\!\!\!\!&\!\!\!\!\!\! $0.0~\%$ \!\!\!\!\!\!&\!\!\!\!\!\! $+9.8\times 10^{-6}$ \!\!\!\!\!\!&\!\!\!\!\!\! $+4.43$ ~~~~~~&\!\!\!\!\!\! $0$   ~~~~~~\\
$-   $   \!\!\!\!\!\!&\!\!\!\!\!\! $5.5~\%$ \!\!\!\!\!\!&\!\!\!\!\!\! $+5.1\times 10^{-6}$ \!\!\!\!\!\!&\!\!\!\!\!\! $+3.59$ ~~~~~~&\!\!\!\!\!\! $2$   ~~~~~~\\
\hline                                                                                                                                                      
$0.03$   \!\!\!\!\!\!&\!\!\!\!\!\! $63 ~\%$ \!\!\!\!\!\!&\!\!\!\!\!\! $-2.7\times 10^{-5}$ \!\!\!\!\!\!&\!\!\!\!\!\! $-1.54$ ~~~~~~&\!\!\!\!\!\! $20$  ~~~~~~\\
$0.06$   \!\!\!\!\!\!&\!\!\!\!\!\! $50 ~\%$ \!\!\!\!\!\!&\!\!\!\!\!\! $-5.5\times 10^{-3}$ \!\!\!\!\!\!&\!\!\!\!\!\! $-0.47 $ ~~~~~~&\!\!\!\!\!\! $549$ ~~~~~~\\
$0.06$   \!\!\!\!\!\!&\!\!\!\!\!\! $67 ~\%$ \!\!\!\!\!\!&\!\!\!\!\!\! $-3.6\times 10^{-5}$ \!\!\!\!\!\!&\!\!\!\!\!\! $-1.90$ ~~~~~~&\!\!\!\!\!\! $20$  ~~~~~~\\
$0.15$   \!\!\!\!\!\!&\!\!\!\!\!\! $50 ~\%$ \!\!\!\!\!\!&\!\!\!\!\!\! $-1.5\times 10^{-4}$ \!\!\!\!\!\!&\!\!\!\!\!\! $-0.55 $ ~~~~~~&\!\!\!\!\!\! $64$  ~~~~~~\\
$0.15$   \!\!\!\!\!\!&\!\!\!\!\!\! $76 ~\%$ \!\!\!\!\!\!&\!\!\!\!\!\! $-5.6\times 10^{-5}$ \!\!\!\!\!\!&\!\!\!\!\!\! $-2.75$ ~~~~~~&\!\!\!\!\!\! $22$  ~~~~~~\\
$0.25$   \!\!\!\!\!\!&\!\!\!\!\!\! $40 ~\%$ \!\!\!\!\!\!&\!\!\!\!\!\! $+1.0\times 10^{-4}$ \!\!\!\!\!\!&\!\!\!\!\!\! $+0.06  $ ~~~~~~&\!\!\!\!\!\! $142$ ~~~~~~\\
$0.25$   \!\!\!\!\!\!&\!\!\!\!\!\! $50 ~\%$ \!\!\!\!\!\!&\!\!\!\!\!\! $-1.0\times 10^{-4}$ \!\!\!\!\!\!&\!\!\!\!\!\! $-0.65 $ ~~~~~~&\!\!\!\!\!\! $45$  ~~~~~~\\
$0.25$   \!\!\!\!\!\!&\!\!\!\!\!\! $60 ~\%$ \!\!\!\!\!\!&\!\!\!\!\!\! $-1.0\times 10^{-4}$ \!\!\!\!\!\!&\!\!\!\!\!\! $-1.37$ ~~~~~~&\!\!\!\!\!\! $32$  ~~~~~~\\
$0.25$   \!\!\!\!\!\!&\!\!\!\!\!\! $70 ~\%$ \!\!\!\!\!\!&\!\!\!\!\!\! $-1.1\times 10^{-4}$ \!\!\!\!\!\!&\!\!\!\!\!\! $-2.14$ ~~~~~~&\!\!\!\!\!\! $29$  ~~~~~~\\
$0.25$   \!\!\!\!\!\!&\!\!\!\!\!\! $80 ~\%$ \!\!\!\!\!\!&\!\!\!\!\!\! $-1.4\times 10^{-4}$ \!\!\!\!\!\!&\!\!\!\!\!\! $-2.99$ ~~~~~~&\!\!\!\!\!\! $30$  ~~~~~~\\
$0.25$   \!\!\!\!\!\!&\!\!\!\!\!\! $84 ~\%$ \!\!\!\!\!\!&\!\!\!\!\!\! $-1.2\times 10^{-4}$ \!\!\!\!\!\!&\!\!\!\!\!\! $-3.40$ ~~~~~~&\!\!\!\!\!\! $30$  ~~~~~~\\
$0.35$   \!\!\!\!\!\!&\!\!\!\!\!\! $50 ~\%$ \!\!\!\!\!\!&\!\!\!\!\!\! $-1.0\times 10^{-4}$ \!\!\!\!\!\!&\!\!\!\!\!\! $-0.70 $ ~~~~~~&\!\!\!\!\!\! $43$  ~~~~~~\\
$0.35$   \!\!\!\!\!\!&\!\!\!\!\!\! $89 ~\%$ \!\!\!\!\!\!&\!\!\!\!\!\! $-1.9\times 10^{-4}$ \!\!\!\!\!\!&\!\!\!\!\!\! $-3.74$ ~~~~~~&\!\!\!\!\!\! $38$  ~~~~~~\\
\hline
\hline
  \end{tabular}
  \end{center} 
  \caption{Summary of the properties of models built with
    prescription~\eqref{eq:fpsi_const} (first two lines) and with the new
    prescription (\ref{Fofpsi}), for $B_p = 10^{15}$~G. }
\label{tab1}
\end{table}


\section[]{Concluding remarks}

We have discussed a novel procedure for obtaining twisted-torus
equilibrium configurations of nonrotating magnetized NSs where the
magnetic field energy in the toroidal component can be as high as the
poloidal one, or even higher. In previous twisted-torus models based on a
simpler prescription for the electric currents, in fact, the toroidal
magnetic energy was at most $\sim 10\%$ of the total. However, with a
suitable choice of the azimuthal currents it is possible to build new
equilibria with toroidal fields containing up to $\sim 90\%$ of the total
internal magnetic energy and toroidal magnetic fields that are on average
about three times larger than the poloidal ones. When compared with the
poloidal-field-dominated geometries proposed in the past, our
configurations represent equally valid candidates for NS interiors and
possibly more realistic ones.

An important implication of our findings is that for a fixed exterior
field strength, stars with larger energy in toroidal-fields have to
have a much larger magnetic energy stored in the interior. As a
result, if NSs have internal magnetic fields close to a twisted-torus
geometry with a high toroidal content, say, $E_{\mathrm{tor}} /
E^{\,\mathrm{int}}_{\mathrm{m}} \gtrsim 10\%$, then the internal magnetic
energy would be higher than commonly assumed, with a potentially strong
impact on the emission properties. As an example, for the new solutions
with $E_{\mathrm{tor}} / E^{\,\mathrm{int}}_{\mathrm{m}} \gtrsim 40\%$,
the GW emission associated with the magnetically induced stellar
deformations is about one order of magnitude larger than for
poloidal-field-dominated configurations.

Finally, the new toroidal-field-dominated configurations could be stable
over several Alfv\'en timescales, in contrast with what known for
poloidal-field-dominated stars. Should this be confirmed by nonlinear
simulations, it would provide strong support to the idea that these
configurations are realistic representations of the stellar
interior. This will be considered in our future work.

\bigskip
\noindent We thank V. Ferrari, K. Glampedakis, J.A. Pons, S.K. Lander
and B. Haskell for useful discussions. RC is supported in part by the
Humboldt Foundation. Support comes also from the DFG grant
SFB/Transregio~7 and by ``CompStar'', a Research Networking Programme
of the European Science Foundation. 


\bibliographystyle{MN2e}
\bibliography{aeireferences}


\label{lastpage}

\end{document}